\documentclass[%
 reprint,
 amsmath,amssymb,
 aps,
]{revtex4-2}

\usepackage{graphicx}% Include figure files
\usepackage{dcolumn}% Align table columns on decimal point
\usepackage{bm}% bold math
\usepackage{hyperref}% add hypertext capabilities
\usepackage{subcaption}
\usepackage{natbib}

\begin{document}

\preprint{APS/123-QED}

\title{On the origin of the gamma-ray burst GRB221009A}%

% \altaffiliation{marcel_oliveira@id.uff.br}%Lines break automatically or can be forced with \\
\author{C. N. Navia}
 \email{carlos\_navia@id.uff.br}
 \author{M. N. de Oliveira}
%\email{marcel\_oliveira@id.uff.br}
 \author{B. O. Felicio}
%\email{marcel\_oliveira@id.uff.br}
\affiliation{
 Instituto de Física, Universidade Federal Fluminense, 24210-346, Niterói, RJ, Brazil}%


\author{A. A. Nepomuseno}
\affiliation{Departamento de Ciências da Natureza, Universidade Federal Fluminense, 28890-000, Rio das Ostras, RJ, Brazil
 }%
%\affiliation{
%\affiliation{%


\date{\today}% It is always \today, today,

\begin{abstract}
We analyze a bright and rare burst, GRB20221009A, showing through NED data that the angular distance
between GRB221009A and the supernova SN 2022xiw is almost the same angular separation with
galactic objects, i.e., the SN-GRB connection at z = 0.151 is not robust. Gamma rays with up to
18 TeV detected in association with the GRB constrain the attenuation to which they are subject
at cosmological distances. GR221009A is out of scale relative to other long-lasting GRBs and comes from a region with an excess of soft gamma repeaters (SGR), suggesting that GRB221009A
is a giant SGR, with energy release Eiso $\sim 10^{44}$ erg. The GRB’s propagation across the galactic plane
reinforces this assumption.
\end{abstract}

\maketitle

%\tableofcontents

\section{\label{sec:introduction}Introduction}

The Fermi Gamma-ray Space Telescope is successful in the detection of
gamma rays from the keV to TeV energy range, such as the detection on
September 14, 2015, of a short GRB (kilonova) triggered by the Fermi
GRM and so far, the unique GRB coincident with a gravitational wave
\cite{conn16}. The discovery of a pulsar from a supernova remnant in
2008 as a purely gamma-ray pulsar \cite{albe06}. Thousand of GRB,
including the GRB 130427A, the ultra-high energy GRB \cite{acke14}. As
well as several hundred solar flares and so on.
The high-energy gamma-ray (1-100 GeV range) detection by Fermi-LAT,
from the direction of the galactic center \cite{ajel16}, suggested
point sources, such as a pulsar population from the galactic bulge.
However, other sources like cosmic ray particles interact with the bulge surrounding the galactic
center \cite{port17}, cannot be discarded.
That is a very active area of research due to the expectation of
gamma-ray excess from annihilating dark matter particles
\cite{gond99,agra14}.

The Fermi GBM detects gamma-ray produced by GRB, solar flares, and
other transient events, such as the SGR, with photon energies from 8 keV to 40 MeV, and a large field of view on the sky with good temporal and
spatial resolution.

On the other hand, we can observe only a low percentage of GRBs that
occur in the “visible” Universe due to emission along two jets in
opposite directions. Only if the Earth is in the path of these jets
it can be detected. GRBs have a typical energy release of
$10^{51}$ erg \cite{frai01}.
That means that if a GRB occurred within the Milky Way, with one of
its jets pointed at Earth, it would be capable of causing great
devastation. Some astrobiological effects of GRBs on the Milky Way are
discussed in \cite{gowa16}.

According to the BATSE GRBs data \cite{kouv93}, GRBs come in two types.
Those with a T90 duration above $\sim$ 2 s, so-called as long, and
those with a T90 duration less than$\sim$ 2 s, so-called as short.

There is some evidence to show that long GRBs originate in the
collapse of massive stars (Hipernovae) in young distant galaxies \cite{mesz06}. Most long GRBs are
followed by an afterglow, allowing us to determine the redshift
(distance) of their host galaxies and the angular width of the two
jets.
The association of short GRBs is possible with
the merging of compact objects in binary systems, such as a double
neutron star (NS), or an NS and a black hole (BH) system
\cite{eich89,nara92,naka07}.

This work was born, by analyzing the Fermi GRM trigger catalog, more
precisely, the excess of triggers around the coordinates RA=19h 38m
9.6s, DEC=+22d 0.1m. The aim was to see if there was any sign of this
excess in the Fermi GRM burst catalog. We found in this region the GRB221009A, the brightest explosion ever observed, also called BOAT (Brightest of All Time) 

GRB221009A is the burst with the highest energy fluence ever observed, and together the enormous release of energy $E_{iso}\sim 10^{55}$ erg considering the GRB at z=0.151, put GRB221009A out of scale when compared to other long-lasting GRBs. 

GRB221009A and the supernova SN 2022xiw appear to be connected \cite{srin23,fult23,ocon23}. However, the absence of heavy elements in the JWST SN spectrum indicates an unusual SN-GRB connection \cite{blan24}. Also, we show other features of GRB221009A that challenge the cosmological origin of this burst, including the wide angular separation between the GRB and SN, considering two objects at z=0.151. The observed gamma-ray counterpart beyond TeV energies in association with the GRB and the GRB221009A's propagation through the galactic plane.
. 

We show that a giant soft gamma-ray repeater originated the GRB221009A and it is more likely than a cosmological origin. Here, we report details of these observations.

\section{Fermi GRM Burst analysis}
\label{analysis}

The BATSE also showed that GRBs come from all over the sky
isotropically, strongly suggesting they are extragalactic explosions.
The first redshift measurements from GRBs by Beppo-SAX confirmed the
GRBs extragalactic origin \cite{anto00}.
However, an isotropy test in the angular distribution of GRBs from
Fermi/GBM is found on \cite{tarn17}. Showing through several
statistical tests that short GRBs are distributed anisotropically in
the sky. Even so, the long ones still have an isotropic distribution.

Here, we analyzed 3721 GRBs in the Fermi GBM Bursts Catalog 
\cite{von20,grub14,von14,bhat16} from 
July 14, 2008, to March 3, 2024. Of them, 619 are short GRBs.
We used twice the LOWESS (Locally Weighted Scatterplot Smoothing)
method \cite{cleve79,nord21}.
The first is to foresee the smooth declination trend curve as a
function of the right ascension, and the second is to obtain a smooth
right ascension trend curve as a function of the declination.

\begin{figure}[h]
\vspace*{-0.0cm}
\hspace*{-0.7cm}
\centering
\includegraphics[width = 4.0 in]{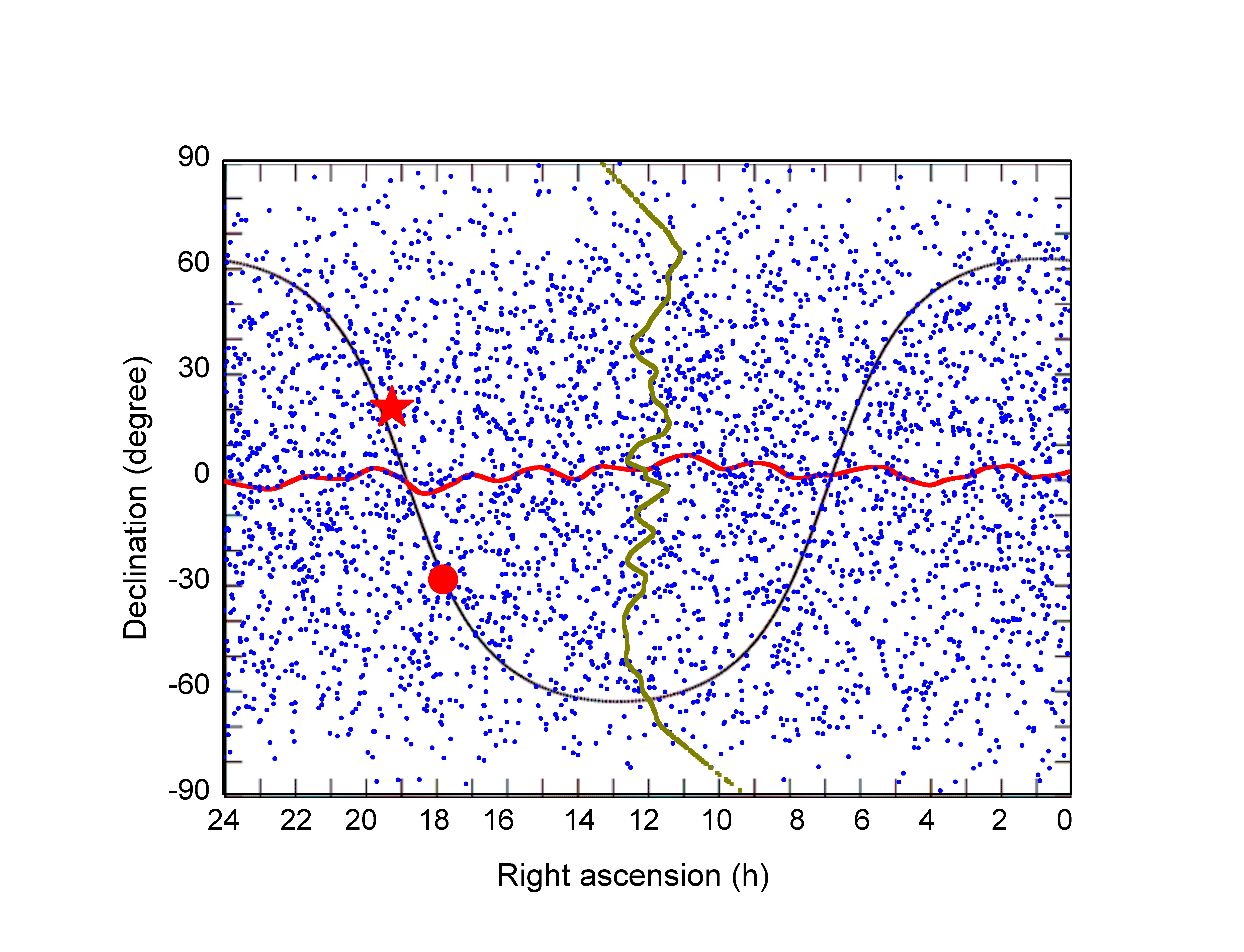}
\vspace*{-0.7cm}
\caption{Scatter plot, using equatorial coordinates (RA, DEC) from the Fermi GRM burst catalog from July 14, 2008, to March 3, 2024 totaling 3721 bursts (blue dots).
 The plot includes the Sagittarius A* coordinates (big red dot), defined as the galactic center (big red dot), the black curve representing the galactic plane, and the red curve (yellow curve) representing the (LOWESS) smooth declination (right ascension) trend curves.
The red star represents the GRB221009A.
}
\label{scatter1}
\end{figure}

Fig.~\ref{scatter1} shows the scatter plot using equatorial coordinates (RA, DEC) of these 3721 Fermi GRM bursts (blue points). The plot includes the Sagittarius A* coordinates, defined as the galactic center (big red dot), the black curve representing the galactic plane, and the red curve (yellow curve) representing the (LOWESS) smooth declination (right ascension) trend curves. The small oscillations (red curve), around a $0.85^{\circ}$ declination and a standard deviation of $2.30^{\circ}$, represent a roughly isotropic distribution of GRBs in the sky.

We want to highlight the big red star in Fig.~\ref{scatter1},
representing an extraordinary, rare, and long-lasting burst, the
GRB221009A. It is almost over the black curve (galactic plane).
GRB 221009A was triggered
and located by Swift-BAT (Dichiara et al.,
\url{https://gcn.nasa.gov/circulars/32632} initially as a new
transient Swift J1913.1+1946 (triggers=1126853 and 1126854).
Fermi GRM also triggered and located the event (Veres et al.,
\url{https://gcn.nasa.gov/circulars/32636}). Indicating in case the
event is a GRB, it is the brightest among the GBM-detected GRBs.
The GBM light curve of the GRB221009A (Lesage et al.,\url{GCN Circular
32642}) consists of two emission episodes, a single isolated peak
followed by a longer, extremely bright, multi-pulsed emission episode
with a duration (T90) of about 327 s (10-1000 keV). The event fluence
(10-1000 keV) from T0+175 to T0+1458 s is on the order of $(2.912 \pm
0.001)\times 10^{-2}$ erg/cm$^2$.
The long-lasting afterglow is a characteristic common feature of the
most violent bursts. So far, none of the violent bursts reached the
fluence of the GRB221009A, which exceeds by almost two orders of
magnitude, constraining the standard models of GRBs \cite{ocon23}.
GRB221009A also was detected by Fermi-LAT (Bissaldi et al.,
GCN 32637), INTEGRAL (SPI-ACS), Konus-Wind, and triangulated by IPN
(DSvinkin et al. 2022, GCN 32641).

%%%%%%%%%%%%%%%%%%%%%%%%%%%%%%%%%%%%%%%%%%%
 
\begin{figure}
\vspace*{-0.0cm}
\hspace*{-0.7cm}
\centering
\includegraphics[width = 4.0 in]{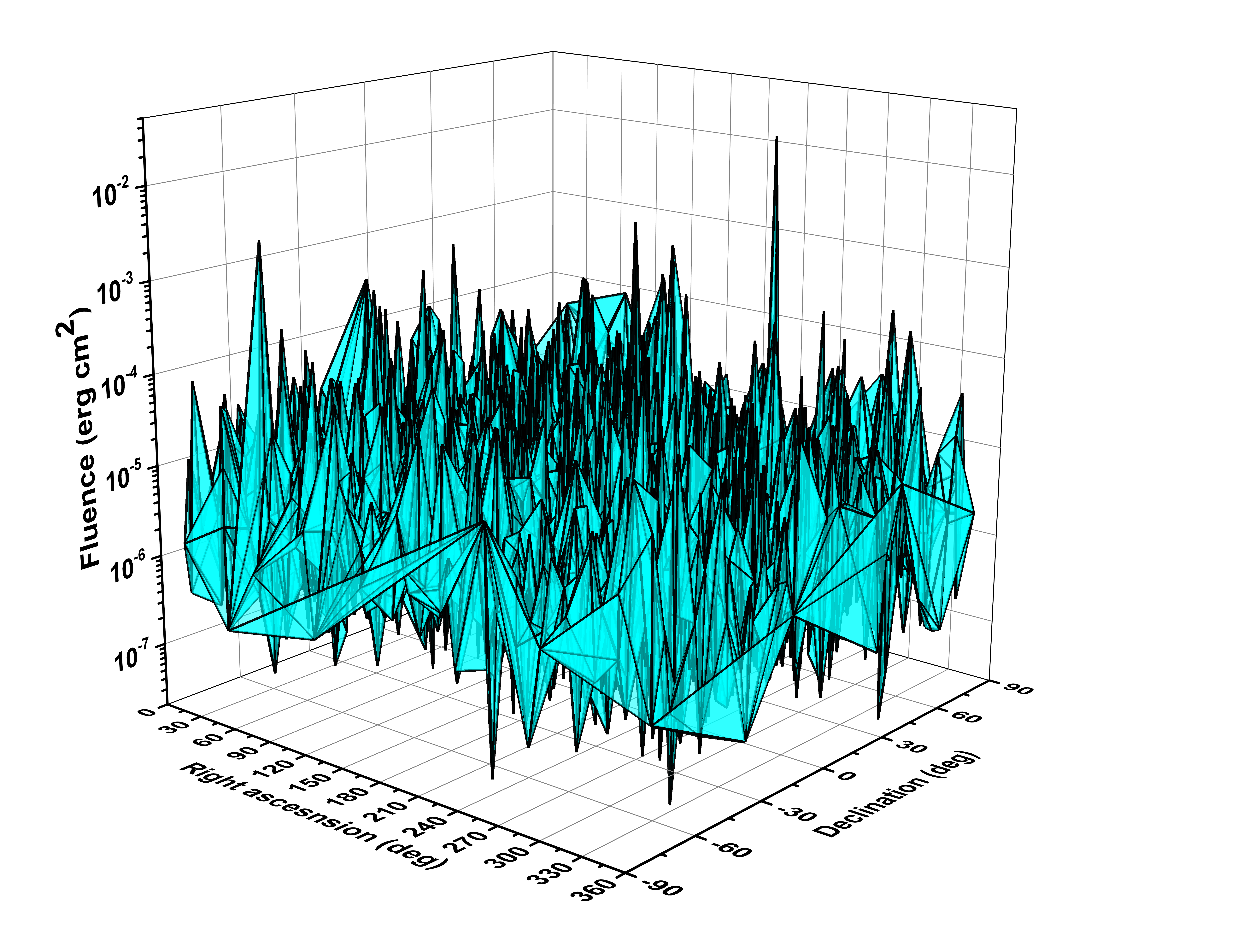}
\vspace*{-0.7cm}
\caption{3D Distribution of the burst's energy fluence from the Fermi GRB Burst Catalog, totaling 3721 bursts.
The highest peak shows the exceptionally long-lasting GRB, the GRB221009A, 
observed with a significance
of 61.0$\sigma$ (fluence 0.039 erg cm$^{-2}$, at coordinates RA=19h 38m 9.6s and DEC=+22d 0.1m. 
}
\label{3D_1}
\end{figure}

Fig.~\ref{3D_1} shows the 3D surface (RA, DEC) plot weighted according
to the bursts energy fluence of the 3721 bursts from the Fermi GRM
catalog.
The plot shows almost an isotropic distribution, with mean
fluence as $(0.2352 \pm 6.4184)\times 10^{-4}$ erg
cm$^{-2}$.
However, the 3D plot also shows the exceptionally rare and long GRB,
the GRB221009A with significance
of 61.0$\sigma$ and coordinates RA=19h 38m 9.6s and DEC=+22d 0.1m.
The burst fluence (0.038833 erg cm$^{-2}$) is, so far, the highest
fluence observed. 

\subsection{SN 2022xiw GRB221009A connection}
\label{connection}

The spectroscopic observation (Ugare Postigo et al.,
\url{https://gcn.nasa.gov/circulars/32800});(Maiorano et al.,
\url{https://gcn.nasa.gov/circulars/32850}) and the optical
counterpart of the GRB 221009A
suggests a connection between the faint supernova SN 2022xiw and the GRB221009A at redshift z=0.151 .
Furthermore, some parameters of the host galaxy extinction constrain
the long-lasting afterglow emission of the GRB, and only the upper
limits are consistent \cite{srin23}. Also,
even considering the connection between SN 2022xiw and the GRB221009A
as plausible, the SN emission appears decoupled from the GRB central
engine.
In addition, there are uncertainties to consider an SN signature in
the extrapolated afterglow of the GRB221009A light curve
\cite{fult23}.
The release of isotropic energy of GRB221009A is exceptionally high
\cite{srin23}, seven orders of magnitude greater than the energy of
the supernova.
Also, the large opening angle of the jets ($\theta\sim$ 0.4 rad)
\cite{ocon23} contrasts with the value inferred from afterglow
observation ($\theta\sim$ 0.01 rad). 

The supernova associated with GRB221009A was also detected by the James Web Space Telescope (JWST) \cite{blan24}, but with no evidence of heavy elements around the exploding star as would be expected. That indicates that the SN-GRB connection is unusual or the SN-GRB connection is uncoupled. All these observations restrict
the SN 2022xiw GRB221009A connection.

\begin{table*}
  \begin{center}
    \caption{Coordinates of the two NED objects with the smallest angular separation of GRB221009A. (Coordinates from \url{https://ned.ipac.caltech.edu/}.)}
    \label{table1}
    \begin{tabular}{ccccc} % <-- Changed to S here.
    %& &  &\textbf{J2000} & \textbf{J2000}  & &\\
    \hline
      Object name & RA & DEC &Type & Separ. \\
      \hline

GRB221009A & 19h 13m  03.8s & +19d 46m 23s &GammaS & \\
\hline

SN 2022xiw & 19h 13m 03.5s & +19d 46m 24s & SN  &0.076 arcmin\\

WISEA J191303.16+194622.6 & 19h 13m 03.2s & +19d 46m 23s & IrS (Galactic)&  0.150 arcmin\\
 \hline 
    \end{tabular}
  \end{center}
\end{table*}

The NED coordinate data (see TABLE I) for the objects SN 2022xiw and
GRB221009A allows us to have the angular separation between them
as $\theta \sim $0.073 arcmin ($2.235\times 10^{-5}$ rad). This angle
is the ratio of arc length to radius length (distance to objects). For
objects at z=0.151 (nearby universe), the relationship between the
Hubble constant and redshift is an indicator of distance as
$D=cz/H_o\sim$ 647 Mpc, where $c$ is the light speed and $H_o\sim 70$ km/s/Mpc is the Hubble constant. Thus, the arc length between these two objects is

\begin{equation}
arc\;length = \theta \times D \sim 0.014\; Mpc.
\end{equation}

This linear size (distance $\sim$0.014 Mpc) between SN 2022xiw and
GRB221009A's origin point is too big, almost the Milky Way diameter 
(0.016 MPC). This result strongly constrains the SN 2022xiw GRB221009A connection.

%According to \cite{gold16}, the spectroscopic redshift distribution %of 335 long GRBs has an average redshift of $z_M \sim 1$. Also, %according to \cite{gold16}, the difference between the redshift %distribution for observed GBM GRBs and the Swift-BAT GRBs (which have %reshifts with a higher average)  can explained by the requirement %that either the Swift-BAT must have observed the GRB or that the GRBs %was bright and seen in the Fermi-LAT.

%Thus, the redshift associated with GRB221009A must be less than %z=0.0076, smaller than the assigning value of z=0.151. Again, this %result constrains the SN 2022xiw GRB221009A connection.

\subsection{Gamma-rays beyond TeV from GRB221009A}
\label{beyond}

The Large High Altitude Air-shower Observatory (LHAASO) in Tibet
claims the detection of more than 140 gamma rays up to 18 TeV from
GRB221009A during 230 to 900 seconds after the trigger \cite{lhaa23}.
However,
the gamma rays detection in the TeV energy region from a cosmological
source with a redshift of z=0.151 sounds strange due to the strong
absorption of (TeV) gamma rays by photons of the extragalactic
background light (EBL).
The attenuation $\tau_{\gamma \gamma}$ of gamma rays for EBL,
depending on gamma-ray energy and redshift as \cite{dwek13}
\begin{equation}
\tau_{\gamma \gamma}(E_{\gamma},z)=-\log \frac{F(E_{\gamma},z)_{obs}} {F(E_{\gamma})_{source}}.
\end{equation}

According to calculations \cite{prim11}, for a source at z=0.1, the
fraction of remaining photons with energies around 1 TeV is only 35\%,
but this fraction falls to 1.5\%
for gamma rays with 10 TeV.
This behavior restricts the
detection of 140 gamma rays up to 18 TeV in association with
GRB221009A emitted at z=0.151.

\subsection{GRB221009A's propagation within the galactic plane}
\label{galactic_plane}

If galaxies are hosts to stars or binary systems that give rise to
GRBs, there is nothing to suggest that this phenomenon cannot happen
in the Milky Way. However, there is a consensus stating that, so far,
only we are detecting GRBs of extragalactic origin because if they
occurred in our galaxy and pointed toward us, the effects would be
catastrophic.
However, if the origin of a GRB is a place diametrically opposite to
our position in the galaxy, the interstellar dust, and the foreground
the galactic plane could shield the Earth. For instance, we do not see all
the luminosity of the galactic center due to the large amount of
interstellar dust.

\begin{figure}[t!]
\vspace*{+0.5cm}
\hspace*{-0.5cm}
\centering
\includegraphics[width = 4.0 in]{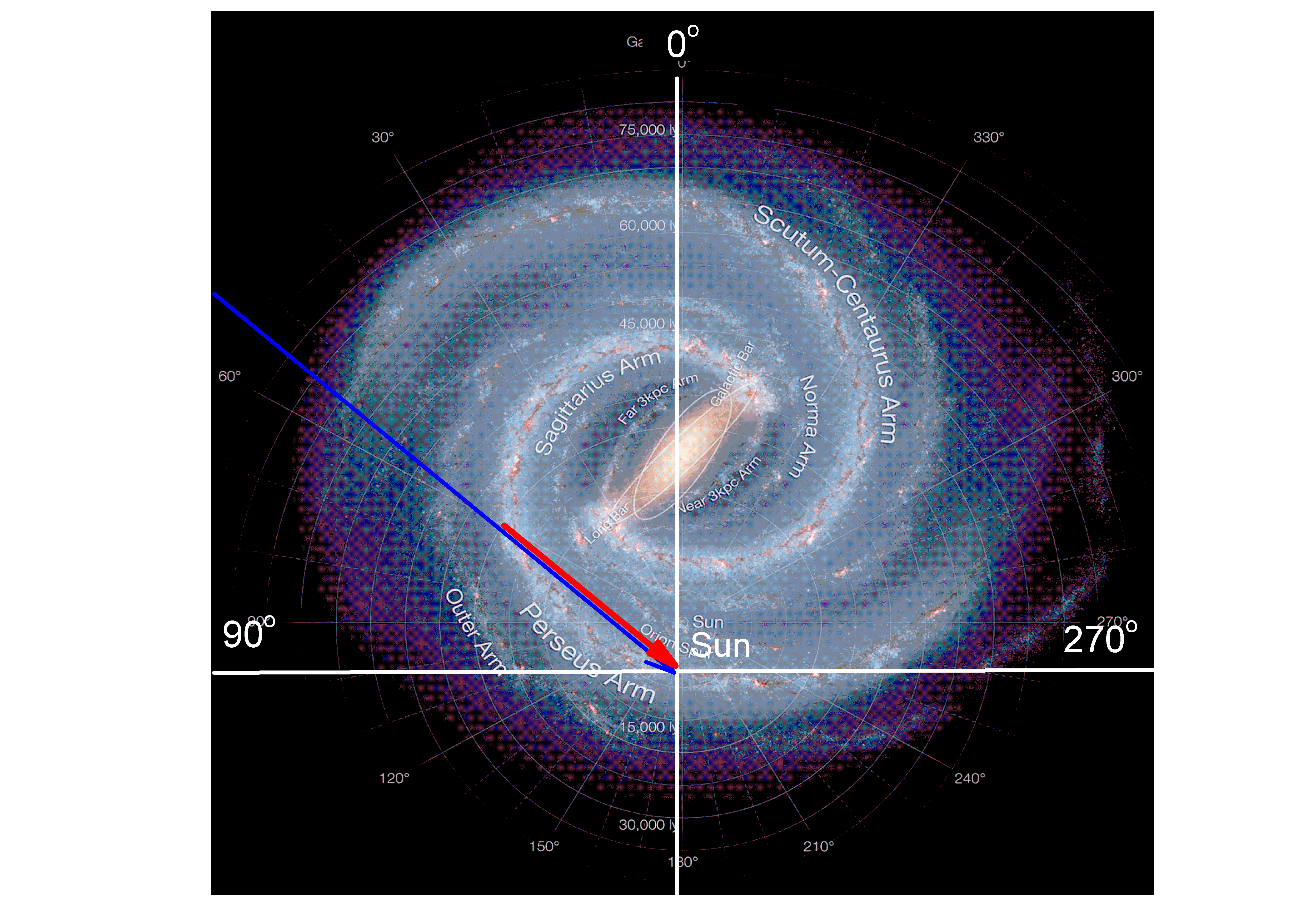}
\vspace*{-0.6cm}
\caption{Sketch of the Milky Way plane and the fundamental plane of the galactic coordinates system centered at the Sun. The blue arrow indicates the GRB221009A arrival direction (longitude l=52.958$^o$,
latitude b=4.32$^o$). These values mean that the burst's propagation was inside the galactic plane. The red arrow represents a probable bust's galactic origin within the Perseus galactic arm.
}
\label{milky_way2}
\end{figure}

Fig.~\ref{milky_way2} shows the plane of the Milky Way, including 
the fundamental plane of the galactic coordinate system centered on the
Sun. The small galactic latitude of the arrival direction (longitude
(l)=52.958$^o$,
latitude (b)=4.32$^o$) of the GRB221009A suggests two alternatives:
a) if the burst is of extragalactic origin, its propagation was across
the galactic plane, as shown by the blue arrow in
Fig.~\ref{milky_way2}. b) If the burst is of galactic origin, its most
likely host would be a star within the galactic Perseus arm or
galactic Outer arm, as shown by the small red arrow in
Fig.~\ref{milky_way2}.

The fact that two objects, the supernova SN 2022xiw (at
z=0.151) and the star (or brown dwarf), WISEA J191303.16+194622.6 (in
the galaxy) have angular separations close to GRB221009A (see TABLE
1) indicates that it is more likely that SN 2022xiw  has no connection to GRB221009A.

\subsection{Fermi Soft Gamma Bursts excesses and GRB221009A} 
\label{sgr}

\begin{figure}
\vspace*{-0.0cm}
\hspace*{-0.7cm}
\centering
\includegraphics[width = 4.0 in]{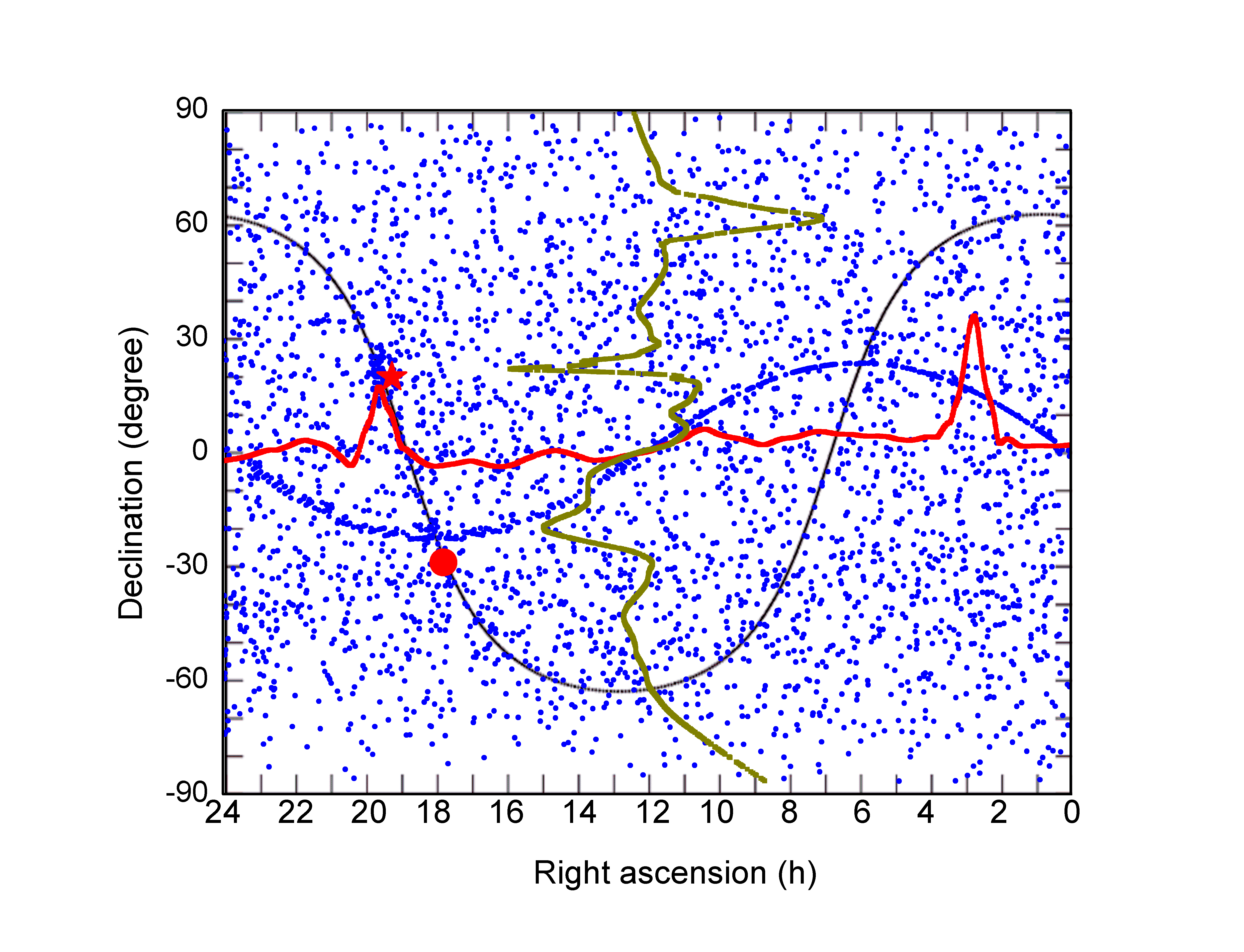}
\vspace*{-0.7cm}
\caption{The same as in Fig.\ref{scatter1}, but considering the Fermi GRM trigger catalog from 2017 to 2023.
}
\label{putoB2}
\end{figure}

\begin{figure}
\vspace*{-0.0cm}
\hspace*{-0.7cm}
\centering
\includegraphics[width = 4.1 in]{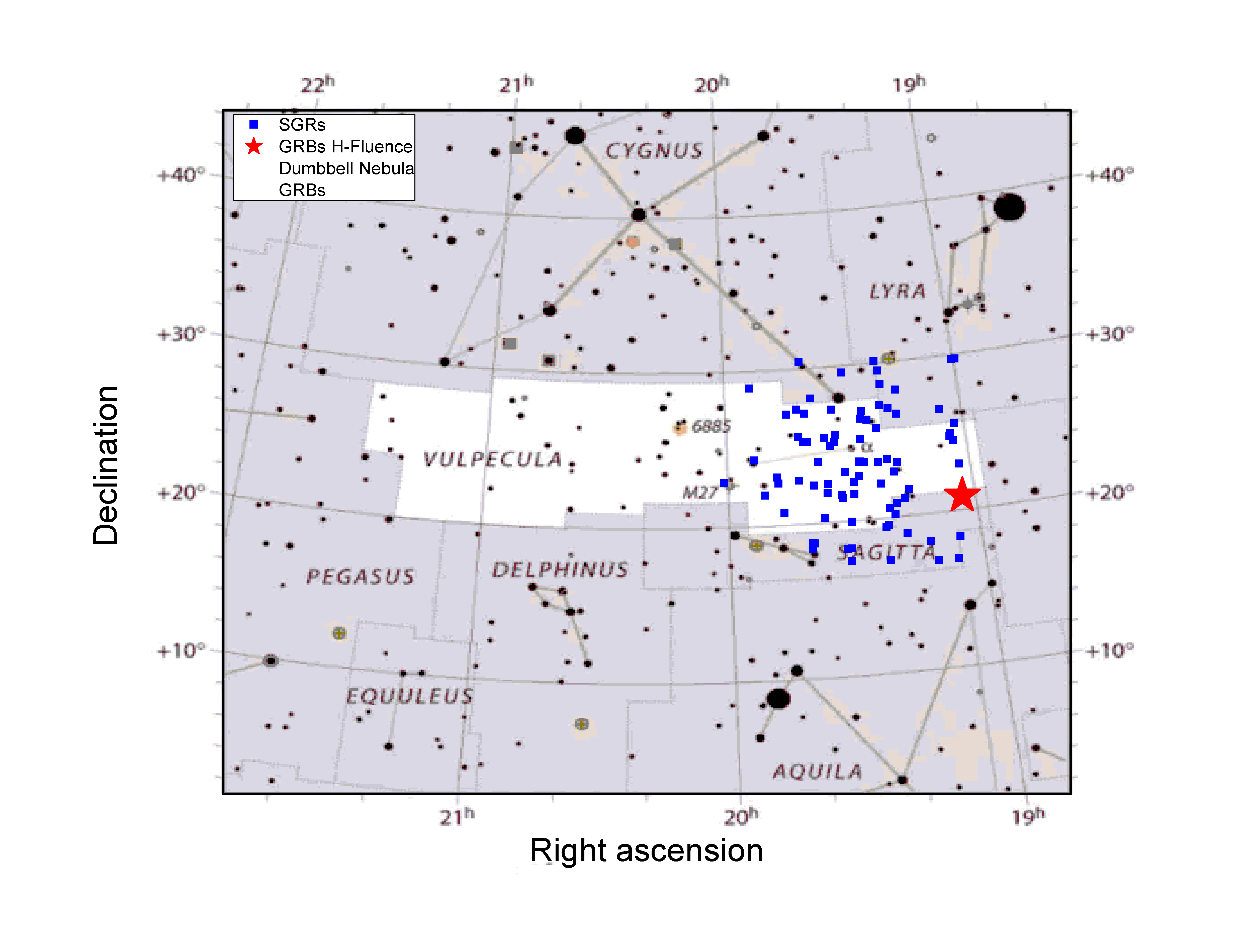}
\vspace*{-0.7cm}
\caption{The figure includes a 1935 square degrees area of the sky region centered on the Vulpecula constellation (268 square degree area) and the scatter plot (RA, DEC) of at least 356 Fermi GRM triggers (excesses) (blue squares) inside a 42.3 square degrees area. Of them, 330 are SGRs, 17 are GRBs, including GRB221009A (red star), and 9 are another type of triggers (Local Particles). 
}
\label{sagitta}
\end{figure}

Fig.~\ref{putoB2} shows the scatter plot using equatorial
coordinates (RA, DEC) of Fermi GRM Trigger Catalog (blue points),
available at
\url{https://heasarc.gsfc.nasa.gov/W3Browse/fermi/fermigtrig.html}
during the last seven years (2017 to 2023).
The symbols
are the same as in Fig.~\ref{scatter1}.
The BRM trigger catalog also includes the GRB221009A, represented by
the red star.
Except for Fig.~\ref{putoB2}, the analysis covers 9853 Fermi GRM
triggers from July 14, 2008, to March 6, 2024.
The origin of these triggers is diverse, but most
of them are GRBs (39\%, including the GRB221009A), followed by solar
flares (SFLARE 21\%),
Terrestrial Gamma-ray Flash (TGF 14\%), Local Particles
(LOCALPAR 13\%), Soft Gamma Repeaters (SGR 6\%), Uncertain (UNCERT
5\%) and Distant Particles 1\%.
From Fig.~\ref{putoB2}, we can see (even with the naked eye) the
existence of an excess of triggers at Ra= 19h 38m 9.6s and DEC= +22d
0.07m, highlighted by the sharp peaks of the two LOWESS curves,
smoothed in declination (red curve) and smoothed in right ascension
(dark yellow curve).
Fig.~\ref{putoB2} strongly suggests that excess triggers
it is on the galactic plane, represented by the black curve.
After zoon in the scatterplot, around the excess triggers and shown in
Fig. ~\ref{sagitta}, we can see an area of the sky with 1935 square
degrees around the Vulpecula constellation (area of 268 square
degrees). The blue squares on the right represent at least 356 excess
triggers within an area $\sim$42.3 $deg^2$. Of these, 330
are SGRs, 17 are GRBs (including GRB221009A, red star), and 9 are
another type of trigger.

The Vulpecula is a faint constellation 
without brilliant stars, several pulsars, such as the PSR B1919+21, a radio pulsar \cite{hewi69}, and also the soft gamma repeater 
SGR 1935+2154, the first galactic SGR emitting radio bursts \cite{chim20}, most 
of the 330 trigger excess comes from this SGR.

The SGRs are sporadic sources of hard X-rays and soft gamma-ray bursts
\cite{hurl08}, with an (isotropic) energy of $\sim 10^{40}$ erg. 
The most accepted origin of SGRs is the activity of a
magnetar, a type of neutron star with an extremely powerful
magnetic field ($10^9$ T to $\sim 10^{11}$ T) originated by a
core-collapse star (supernova).

\begin{figure}
\vspace*{-0.0cm}
\hspace*{-0.7cm}
\centering
\includegraphics[width = 4.0 in]{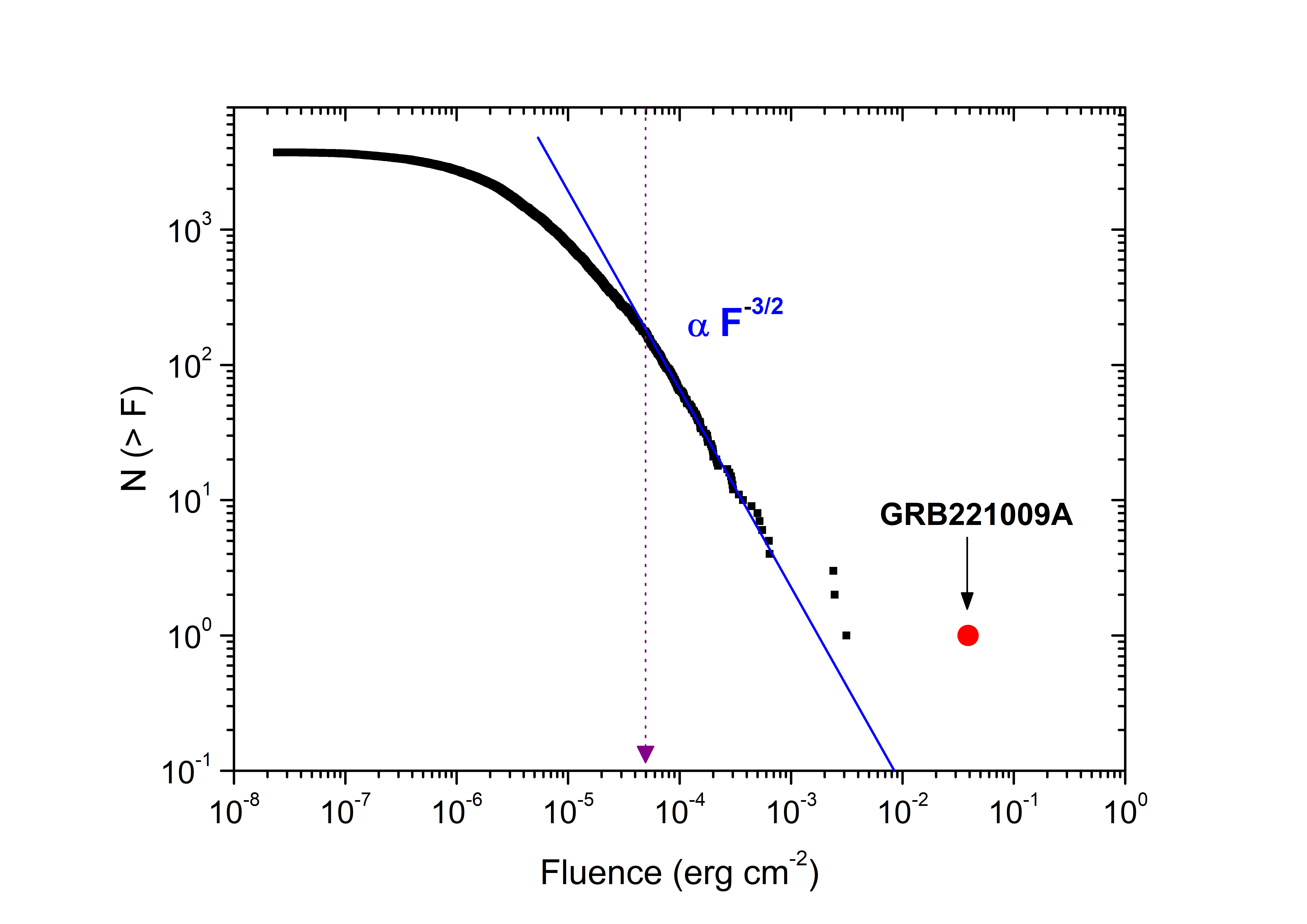}
\vspace*{-0.7cm}
\caption{Distribution of the burst's energy fluence from the Fermi GRM Burst Catalog totalizing 3721 bursts.
The fit for bursts with fluence above $5\times 10^{-5}$ erg cm$^{-2}$ is consistent with a power law, with index $-3/2$. The big red dot indicates the GRB221009A and falls away from the distribution.
}
\label{flue_dis}
\end{figure}

We want to highlight that all 17 GRBs from the excess triggers (from
Fermi GRM trigger catalog) are included in the energy fluence
distribution shown in Fig.~\ref{flue_dis}. If we consider only those
with fluence above 5.0$\times 10^{-5}$ erg/cm$^2$, the distribution is
consistent with a power law fit as $f^{-\alpha}$ with $\alpha = 3/2$.
The value of $\alpha$ means a homogeneous GRB population in Euclidean
space, signifying an extragalactic origin. However,
GRB221009A (big red dot) is outside of this distribution.

We also want to highlight that the extraordinary and rare burst GRB221009A, is within a very active area of magnetars, triggering SGRs, and suggests a giant SGR, as the
origin of GRB221009A.

%%%%%%%%%%%%%%%%%%%%%%%%%%%%%%%%%%%%%%%%%%%%%%%%%%%%%%%%%%%%%%%%%%%%%%%%%%%%%%%%%%%%%%%%%%%%%%%%%%

\begin{figure}
\vspace*{-0.7cm}
\hspace*{-0.5cm}
\centering
\includegraphics[width = 3.9 in]{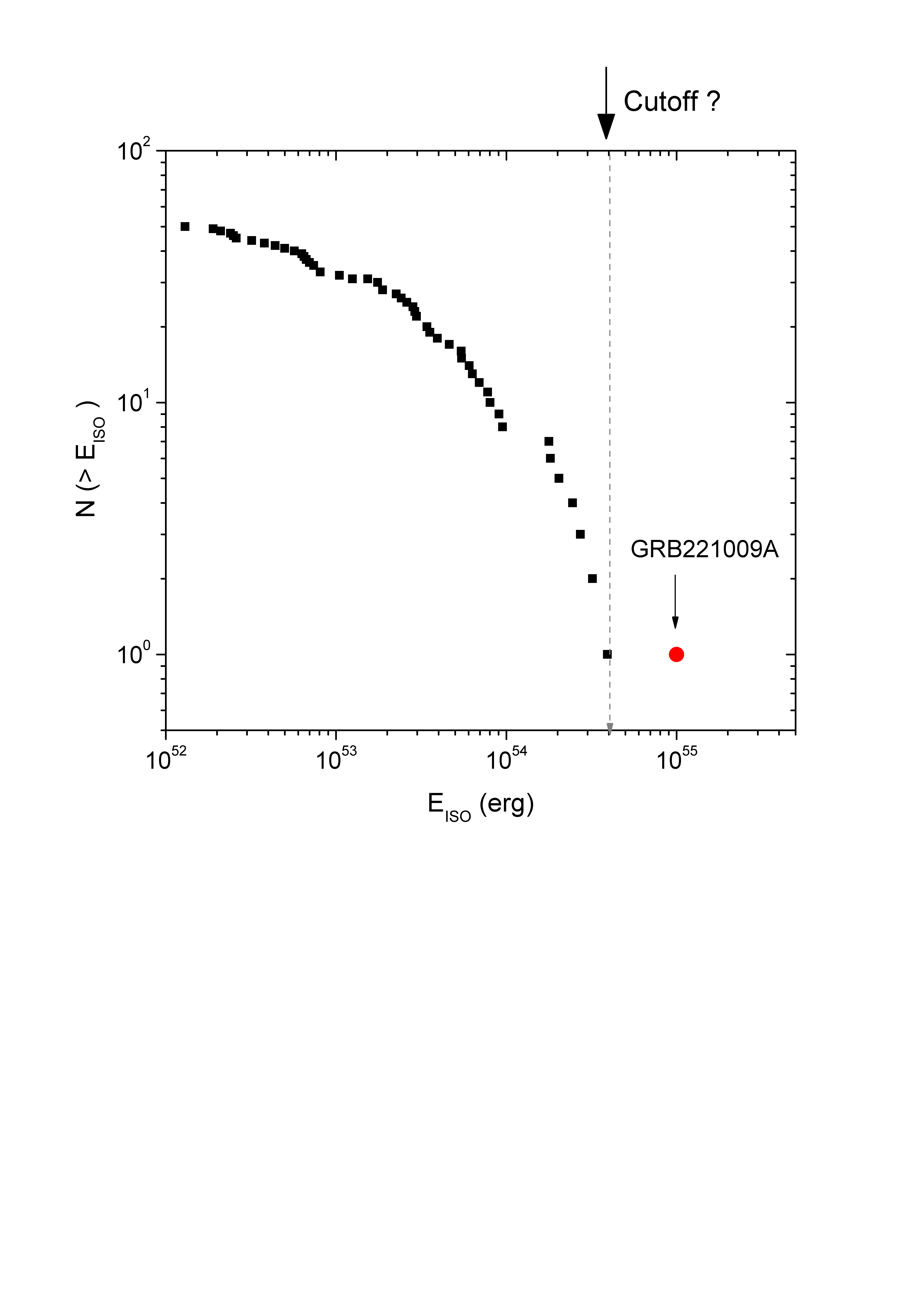}
\vspace*{-5.7cm}
\caption{Distribution of the burst's $E_{iso}$ energy from 55 long Fermi GRM bursts with redshifts in the range of z-1-4 (data from \cite{atte17}). 
The vertical gray line indicates a $E_{iso}$ value cutoff.  The red dot indicates the GRB221009A and falls away from the distribution.
}
\label{eiso_dis}
\end{figure}

\subsection{$E_{iso}$ value: GRB versus SGR}
\label{iso}

The isotropic equivalent energy ($E_{iso}$) is significant to obtain the gamma-ray energy release exhibited by GRBs and the universe's evolution effect \cite{lloy02}. For a nearby universe, $E_{iso}$ and redshift (z) are connected through the relationship \cite{wu12}
\begin{equation}
E_{iso}=4\pi F \frac{D^2(z)}{(1+z)^{1,8}},
\end{equation}
where $F$ is the burst's energy fluence.

Here, we compare the $E_{iso}$ order of magnitude for the GRB221009A with other values from the literature, considering two alternatives regarding the origin of GRB221009A.

(a) GRB221009A is a GRB of cosmological origin
(z=0.151). That means that its origin occurred at a distance
D$\sim$ 647 Mpc, giving $E_{iso} \sim 1.0\times 10^{55}$ erg \cite{ocon23}.
This value is above to the sharp cutoff point above $\sim 4\times 10^{54}$ erg in the $E{iso}$ distribution \cite{atte17}.

Fig.~\ref{eiso_dis} shows the $E_{iso}$ distribution of 55 long Fermi GRM bursts with redshift in the range of z=1-4, the data was compiled from \cite{atte17}. The red dot represents the GRB $E_{iso}$ energy of GRB221009A under the assumption of a cosmological origin and would be the first GRB to break this cutoff.

Also, this enormous energy value of $E_{iso}$ is seven orders of magnitude greater than the energy of its associated supernova, SN 2022xiw \cite{srin23}.

(b) GRB221009A is a SGR of galactic origin
(z=0), this means that its origin occurred at a distance
D$\sim$ 0.008 Mpc (radii of the Milky Way), giving $E_{iso} \sim 1.1\times 10^{44}$ erg.
This value is close to that found in the literature for giant soft gamma repeaters \cite{mere08}, such as SGR 0526-66
($E_{iso}\sim 3.6\times 10^{44}$ erg) and SGR 1900+14
($E_{iso}\sim 1.2\times 10^{44}$ erg).

A magnetar, producing a giant SGR at the Perseus galactic arm at a distance   $\sim$0.008 Mpc from Earth, explains why GRB221009A appears out of scale relative to other long-lasting GRBs. Also, explains
the intense brightness of GRB221009A observed in several detectors around Earth.

\section{Discussions and conclusions}
\label{conclusion}

In this article, we describe several features concerning the
exceptionally bright and rare GRB221009A that challenge a cosmological
origin (z=0.151) for the GRB and its connection with the supernova SN
2022xiw.

(a) The angular separation between GRB221009A and SN 2022xiw is only half of the angular separation 
between the GRB and an object within the galaxy, the WISEA
J191303.16+194622.6 (see TABLE I).
This behavior indicates two possibilities: GRB and SN are galactic
objects or SN at z=0.151 without any connection with the GRB.

(b) The supposed connection of GRB221009A to SN 2022xiw is not robust. The large opening angle of the jets contrasts with the small value inferred from afterglow observation  \cite{ocon23}. The absence of heavy elements in the JWST SN spectrum indicates an unusual SN-GRB connection \cite{blan24}.

(c)The fast increase in attenuation of high-energy gamma rays as the
energy and distance increase constrains the observation of extragalactic sources emitting gamma
rays beyond the TeV energies.

The detection of a large number of gamma rays beyond TeV energies (up
to 13 TeV) from GRB221009A by LHAASO
suggests that intergalactic space is not opaque to this type of
radiation than previously expected or that the radiation source is
not at cosmological distances.

(d) The small galactic latitude (b=4.21 degrees) in the arrival direction of GRB221009A means that the GRB propagation was across the galactic plane.
However, the propagation across the galactic plane also suggests
 a galactic origin for the GRB221009A, with a host star (magnetar) at the Perseus galactic arm, or Outer galactic arms (see Fig~\ref{milky_way2}).

(e) GRB221009A is in a region on the galactic plane with
excess Fermi GRM triggers. Most of them are identified as SGRs,
suggesting that GRB221009A is not a GRB but a giant SGR. The energy
release $E_{iso}$ obtained under this assumption is in the same order
of magnitude as the giant SGRs of the literature ($E_{iso}\sim
10^{44}$ erg).

(f) The GR221009A burst occurred out of scale compared to other 
long-lasting bursts. GRBs with fluence above $5\times 10^{-5}$ erg cm$^{-2}$ from the Fermi GRM burst catalog is consistent with a power-law homogeneous
distribution (see Fig.~\ref{flue_dis}), meaning an extragalactic origin to the GRBs. However, GRB221009A falls away from this distribution.

Furthermore, analyzing long GRGs from Fermi/GBM and Konus-Wind in
the range of redshifts z=1–4, shown the existence of a sharp
cutoff of the $E{iso}$ distribution of GRBs around
$\sim 4\times 10^{54}$ erg \cite{atte17}. We have reproduced
this distribution for the Fermi GRM bursts (see Fig.~\ref{eiso_dis}).
Again, the GRB221009A falls away from the distribution. The GRB221009A
is the first GRB to break this cutoff.

The enormous energy release $E_{iso}\sim 10^{55}$ erg obtained for
GRB221009A \cite{ocon23} under the assumption of a cosmological origin
is hard to consolidate with standard GRB models.

From these considerations, we have shown most likely that GRB221009A is a giant SGR. 
A magnetar, producing a giant SGR, probably occurred in the Perseus galactic arm at a distance   $\sim$0.008 Mpc from Earth, which explains the intense brightness observed in several detectors around Earth.

\begin{acknowledgments}

We thank the Fermi Gamma-ray Space Telescope and Gamma-ray Coordinate Network (GCN) for their valuable information and open data policy.
 This work has the support of the Rio de Janeiro State Research Support Foundation (FAPERJ) under grant E 26/010.101128/2018.
\end{acknowledgments}

\bibliography{bibi}% Produces the bibliography via BibTe

\end{document}